\documentclass[english]{article}
\usepackage[T1]{fontenc}
\usepackage[latin1]{inputenc}

\makeatletter

 \newcommand{\lyxaddress}[1]{
   \par {\raggedright #1
   \vspace{1.4em}
   \noindent\par}
 }

\date{}

\usepackage{babel}
\makeatother
\begin{document}

\title{Second virial coefficient in one dimension, \\ 
as a function of asymptotic quantities}

\author{A. Amaya-Tapia$^{1}$, S. Y. Larsen$^{2}$and M. Lassaut$^{3}$}

\maketitle

\lyxaddress{$^{1}$Centro de Ciencias F\'{i}sicas, UNAM AP 48-3, 
Cuernavaca, Mor. 62251, M\'{e}xico.$^{2}$Department of Physics, Temple 
University, Philadelphia, PA 19122, USA.$^{3}$Groupe de Physique
Th\'{e}orique, Institut de Physique
Nucl\'{e}aire, F-91406 Orsay Cedex, France.}

\begin{abstract}
A result from Dodd and Gibbs\cite{dodd} for the second virial coefficient
of particles in 1 dimension, subject to delta-function interactions,
has been obtained by direct integration of the wave functions.
It is shown that this result can be obtained from a phase
shift formalism, if one also includes the contribution of oscillating terms.
The result is important in work to follow, for the third virial coefficient,
for which a similar formalism is being developed.
We examine a number of fine points in the quantum mechanical formalisms.
\end{abstract}
\section*{Introduction}

The quantum mechanical formulation of virial coefficients, or the 
calculation of corrections to classical results, continues to be of concern.
We note, for example, the fairly recent article by Hussien and 
Yahia \cite{hussien}, on the quantum corrections to the classical fourth
virial coefficient for square well potentials. 
Our efforts are, instead, directed to the continuing development of a
fully quantum mechanical formalism, applicable to the higher virials, 
and especially useful at low temperatures.

We recall that, many years ago, Dodd and Gibbs\cite{dodd} were able to
evaluate the
second and third virial of identical particles in one dimension, subject
to repulsive delta function interactions. To do this, they introduced
a complete set of energy eigenfunctions in the traces, and then integrated
to obtain answers in terms of single and two-dimensional integrals.

Their original aim, as is clear from their paper and from Gibbs' thesis,
was to obtain a formalism in terms of the S-matrix, following the
formalism of Dashen, Ma and Bernstein\cite{dashen}, in an effort
to obtain the virials in terms of scattering quantities. This was
not possible, as the authors determined that the `formal limiting
processes are not valid for the singular amplitudes of this system'.

An alternative formalism, proposed by Larsen and Mascheroni\cite{masche},
aimed at expressing the higher virials in terms of eigenphase shifts
and bound state energies, and, if necessary, other on-shell properties,
in a hyperspherical harmonic formalism. Under some constraints, one
is then able to recover the classical results for the 3rd virial,
through a semi-classical approximation\cite{larsen2}. Low temperature
calculations have been carried out for repulsive potentials in 2
dimensions\cite{jei},
and a more sophisticated formalism, involving a hyperspherical adiabatic
basis, has been presented\cite{larsen},

Our aim, in this and subsequent papers, is to use the excellent results
of Dodd and Gibbs to confirm our virial formalism for the second and
third virials, for this particular dimension and interaction, for
which we have analytical and numerical results\cite{gibson,popiel,amaya}.
More specifically, in this short note, we will show that, for the
second virial, the result of Dodd and Gibbs can be rewritten as an
integral, over the energy, of a Boltzmann factor together with the
phase shift - the usual second virial type of result - plus that of
an extra `oscillating' type of term, particular to the contribution of 
even parity wave functions in one dimension. 
This situation is understood\cite{gibson2},
though not necessarily widely, and we discuss it for this particular
case and in greater generality.

\newpage
\section*{The Second Virial Coefficient}

\noindent Following ref. \cite{larsen}, in this section we present
the main steps leading us from the grand partition function to the
expression of the second virial coefficient in terms of the asymptotic
properties of the system. Instead of using the Ursell expansion, we simply
take, directly, the logarithm of the grand partition function.

The latter can be written as

\begin{equation}
\mathit{Q}={\displaystyle \sum_{n=0}^{\infty}z^{n}{}Tr\left(e^{-\beta H_{n}}\right)}=1+zTr\left(e^{-\beta T_{1}}\right)+z^{2}{}Tr\left(e^{-\beta H_{2}}\right)+\cdots\label{part}\end{equation}
 The fugacity $z$ equals $\exp\left(\mu/\mathrm{k}T\right)$;$\;\beta=1/\mathrm{k}T$,
where $\mu$, $\mathrm{k}$ and $T$ are the Gibbs' function per particle, 
Boltzmann's 
constant and the temperature, respectively. $H_{n}$ and $T_{n}$ are
the n-particle Hamiltonian and kinetic energy operators.

It is important to note that

\[
Tr^{}\left(e^{-\beta H_{n}}\right)\rightarrow V^{n}\;\; as\;
 V\;\rightarrow large.\]
 However, the logarithm of the partition function

\begin{flushleft}\[
\begin{array}{ccc}
ln\mathit{Q} & = & ln\left(1+\left[zTr\left(e^{-\beta T_{1}}\right)+z^{2}\, Tr\left(e^{-\beta H_{2}}\right)+\cdots\right]\right)\\
 & = & \left[zTr\left(e^{-\beta T_{1}}\right)+z^{2}{}Tr\left(e^{-\beta H_{2}}\right)+\cdots\right]\\
 &  & -\frac{1}{2}\left[zTr\left(e^{-\beta T_{1}}\right)+z^{2}{}Tr\left(e^{-\beta H_{2}}\right)+\cdots\right]^{2}\end{array}\]
\end{flushleft}

\begin{flushleft}or, equivalently,\end{flushleft}

\begin{equation}
\begin{array}{cl}
ln\mathit{Q} & =z\, Tr\left(e^{-\beta T_{1}}\right)\\
 & +z^{2}\left[Tr\left(e^{-\beta H_{2}}\right)-\frac{1}{2}\left[Tr\left(e^{-\beta T_{1}}\right)\right]^{2}\right]\\
 & +z^{3}\left[Tr\left(e^{-\beta H_{3}}\right)-Tr\left(e^{-\beta T_{1}}
\right)Tr\left(e^{-\beta H_{2}}\right)+\frac{1}{3}\left[Tr\left(e^{-\beta
 T_{1}}\right)\right]^{3}\right] \\
 & +\cdots\end{array}\label{lnq}\end{equation}
 when divided by V, gives coefficients which are independent
of the volume, when the latter becomes large; we call them $b_{l}$.
We can then write for the pressure and the density\begin{equation}
\begin{array}{l}
\frac{p}{\mathrm{k}T}=\frac{1}{V}\, ln\mathit{Q}=\sum_{n=1}^{\infty}b_{n}z^{n}\\
\rho=\frac{N}{V}=z\,\frac{\partial}{\partial z}\left(\frac{1}{V}\, ln\mathit{Q}\right)=\sum_{n=1}^{\infty}n\, b_{n}z^{n}\end{array}\label{eqst}\end{equation}
 where

\begin{equation}
b_{1}=\frac{1}{V}\, Tr\left(e^{-\beta T_{1}}\right)\:;\: b_{2}=\frac{1}{V}\,\left\{ Tr\left(e^{-\beta H_{2}}\right)-\frac{1}{2}\left[Tr\left(e^{-\beta T_{1}}\right)\right]^{2}\right\} ,\; etc.\label{b}\end{equation}
 Elimination of $z$ from Eqs. $\left(\ref{eqst}\right)$ gives the
virial expansion of the equation of state\[
\frac{p}{\mathrm{k}T}=\sum_{n=1}^{\infty}B_{n}\rho^{n}.\]

Let us evaluate the fugacity coefficients.

\begin{equation}
\begin{array}{cl}
b_{1} & =\frac{1}{V}\,{\displaystyle \sum_{k}}<k|e^{-\left(\hbar{}^{2}/2m\right)\beta k^{2}}|k>\\
 & =\frac{1}{V}\frac{1}{(2 \pi)^{3}}\,\int d\overrightarrow{r}\int d\overrightarrow{k}\, 
e^{-i\overrightarrow{k}.\overrightarrow{r}}\, e^{-\left(\hbar{}^{2}/2m\right)
\beta k^{2}}\, e^{i\overrightarrow{k}.\overrightarrow{r}}\\
 & =\frac{1}{\left(h^{2}/2\pi m\mathrm{k}T\right)^{3/2}}
=\frac{1}{\lambda_{T}^{3}},\end{array}\label{b1}\end{equation}
 where we introduced a complete basis set, for positive energies 
in the continuum, and wrote the result in terms of the thermal
wave length, $\lambda_{T}.$

Now, let us to evaluate the second coefficient in three dimensions. 
Consider the Hamiltonian
$H_{2}$, separated in a CM part and a relative one,
 $H_{2}=H_{2}^{cm}+H_{2}^{rel}$.
Note that for the center of mass $m\,\rightarrow2m$, 
thus $\lambda_{T}^{-3}\,\rightarrow2^{3/2}\,\lambda_{T}^{-3}$.

The trace of the CM part becomes

\begin{equation}
\frac{1}{V}Tr\left(e^{-\beta H_{2}^{cm}}\right)=\frac{1}{V}Tr\left(e^{-\beta
 T_{2}^{cm}}\right)=\frac{2^{3/2}}{\lambda_{T}^{3}}\label{b2cm}\end{equation}
 and the trace of the relative part, in the symmetric case, can be
written as

\begin{eqnarray}
Tr^{s}\left(e^{-\beta H_{2}^{rel}}\right) & ={\displaystyle \sum_{m}}^{s}<m|e^{-\beta H_{2}^{rel}}|m>
+\int dq<q|e^{-\beta H_{2}^{rel}}|q> \nonumber       \\
 & ={\displaystyle \sum_{m}}^{s}e^{-\beta E_{m}^{bd}}+\int d\overrightarrow{r}
\int dq\:\psi_{q}^{s*}\left(\overrightarrow{r}\right)\psi_{q}^{s}
\left(\overrightarrow{r}\right)e^{-\beta E_{q}}
\label{trace}
\end{eqnarray}
 where we assumed a spherically symmetric potential and introduced a complete
set of symmetric eigenstates. The
index $q$ includes the continuum variable $k$, as well as the discrete
angular momentum indices $\ell$ and $m$. 
The integration over $\overrightarrow{r}$ is carried out over the reduced range (half) of 
$\overrightarrow{\omega}$, since this then yields all of the configurations of the symmetrical wave
functions. These wave functions are normalized over this reduced range. 
 Our result reads:

\begin{equation}
 {\displaystyle \sum_{m}}^{s}e^{-\beta E_{m}^{bd}}
  +{\displaystyle \sum_{\ell \,= \, even}}
\left(2\ell+1\right)\: \int^{\infty}_{0} dk \, e^{-\beta E_{k} }
\int^{\infty}_{0} dr\,\chi^{*}_{k\ell}
\left(r\right)\chi_{k\ell}
\left(r\right) \ .
\label{tr}
\end{equation}

A method, that has been used to evaluate the integral over $k$,
is to put the system in a box of volume V (see, for instance, ref.
\cite{masche}). The energies will then be discrete, characterized by an 
index $n$, and the eigenfunctions
will be normalized so that their integral over the $r$ gives 1.
 For large
volumes, we can again proceed from the sum to an integral in which,
to estimate the density of states $\frac{dn}{dk}$, we use the asymptotic
form of the wave function \[
\chi_{k\ell}\left(r\right){}_{r_{large}} \sim \sin\left(kr-\frac{\ell\pi}{2}
+ \delta_{\ell}\left(k\right)\right).\]
We require that it go to zero at the boundary of a large sphere, of radius R,
and that therefore the argument equal $n \pi$.
It then follows that

\[
\frac{dn}{dk}=\frac{1}{\pi}\left(R+\frac{d}{dk}\delta_{\ell}\left(k\right)\right)\]
 and that  the continuum contribution equals
\[
Tr^{s}\left(e^{-\beta H_{2}^{rel}}-e^{-\beta T_{2}^{rel}}\right)=\\
{\displaystyle \sum_{\ell \,= \, even}}\left(2\ell+1\right)\:\frac{1}{\pi}
\int^{\infty}_{0} dk\,
\left(\frac{d}{dk}\delta_{\ell}\left(k\right)\right)e^{-\beta E_{k}}.
\]

The difference $b_{2}^{s}-b_{2}^{s0}$ in the second fugacity coefficients
is then
\begin{equation}
{\displaystyle \sum_{m}}^{s}e^{-\beta E_{m}^{bd}}+\frac{2^{3/2}}
{\lambda_{T}^{3}}\,{\displaystyle \sum_{\ell \,= \, even}}
\left(2\ell+1\right)\:\frac{1}{\pi}\int^{\infty}_{0} dk\,
\left(\frac{d}{dk}\delta_{\ell} \left(k\right)\right)
e^{-\beta (\frac{\hbar^{2}}{m})k^{2}}
\label{3dif}
\end{equation}
 and the virial coefficient takes the well known form \cite{boyd,uhlenbeck}
in terms of the energies of the bound states and the derivative of
the eigenphase shifts:
 \begin{equation}
\begin{array}{rl}
B_2\left(T\right)-B_2^{0}\left(T\right)=
 & -\texttt{N}\frac{b_{2}^{s}-b_{2}^{s0}}{\left(b_{1}\right)^{2}}\\
= & -\texttt{N}2^{3/2}\lambda_{T}^{3}\,\left[{\displaystyle \sum_{m}}
^{s}e^{-\beta E_{m}^{bd}}\right.\\
 & \left.+\frac{1}{\pi}{\displaystyle \sum_{\ell \,= \, even}}
\left(2\ell+1\right)\:\int^{\infty}_{0} dk\,\left(\frac{d}{dk}\delta_{\ell}
\left(k\right)\right)
e^{-\beta (\frac{\hbar^{2}}{m})k^{2}}\right] \ . \end{array}
\label{vir}
\end{equation}
 An integration by parts can change the integral to involve the phase shifts
directly.

\section*{Oscillatory Terms}

The passage from the sum, in the box approximation, to the integral,
in the continuum situation, is usually perfectly justified. The correction
terms are negligible, except under some very specific circumstances,
as we shall see.

To investigate this, we follow a method\cite{poll} devised for the
second virial coefficient for anisotropic interactions, where it becomes
impossible, in a box, to simultaneously satisfy a common boundary condition
for all the amplitudes (belonging to different orbital angular momenta) of
any given eigenfunction. It became useful then, and elegant, to directly
evaluate the integral of the square of the wave functions that already 
appears (\ref{tr}) in our formulation for the symmetric case.
To do so requires a simple 'Green's function trick'.

The Schr\"{o}dinger equation for the radial wave function corresponding
to $k$ is

\begin{equation}
-\frac{d^{2}}{dr^{2}}\chi_{k\ell}\left(r\right)+\frac{\ell\left(\ell+1\right)}
{r^{2}}\chi_{k\ell}\left(r\right)+\frac{\hbar^{2}}{2m}\, 
V\left(r\right)\chi_{k\ell}\left(r\right)=k^{2}\,\chi_{k\ell}\left(r\right) \ .
\label{A2}\end{equation}
 Multiplying Eq. (\ref{A2}) by $\chi^{*}_{k'\ell}\left(r\right)$, and the
complex conjugate of the
corresponding equation for k' by $\chi_{k\ell}\left(r\right)$, and then taking
the difference between them, we obtain

\begin{equation}
-\frac{1}{\left(k'^{2}-k^{2}\right)}\,\frac{d}{dr}\left[\chi_{k\ell}
\left(r\right)\,\frac{d}{dr}\chi^{*}_{k'\ell}\left(r\right)-\chi^{*}_{k'\ell}
\left(r\right)\,\frac{d}{dr}\chi_{k\ell}\left(r\right)\right]=
\chi^{*}_{k'\ell}\left(r\right)\chi_{k\ell}\left(r\right) \ .
\label{A3}\end{equation}
 Integrating over the interval $0<r<r_{max}$, using the boundary
conditions at the origin $\chi_{k\ell}\left(0\right)=0$,
\, $\chi_{k'\ell}\left(0\right)=0$
and L'Hospital rule, expression (\ref{A3}) becomes
\begin{equation}
\frac{1}{2k}\left\{ \,\left(\frac{\partial}{\partial k}\chi_{k\ell}
\left(r\right)\right)\,\frac{\partial}{\partial r}\chi^{*}_{k\ell}
\left(r\right)-\chi^{*}_{k\ell}\left(r\right)\frac{\partial^{2}}
{\partial k\partial r}\chi_{k\ell}\left(r\right)\right\}_{r=r_{max}}
 =\int_{0}^{r_{max}}dr\,\chi^{*}_{k\ell}\left(r\right) 
 \chi_{k\ell}\left(r\right) 
\label{A4}
\end{equation}
when $k' \to k$. For $r_{max}$ large enough, we can use the asymptotic
form of the wave function 
\begin{equation}
\sqrt{\frac{2}{\pi}}\sin\left(kr-\ell\pi/2+\delta_{\ell}\left(k\right)\right)
\label{A5}
\end{equation}
 to evaluate the left hand side of the Eq. $\left(\ref{A4}\right)$ as
\begin{equation}
\begin{array}{l}
\frac{1}{\pi}\left(r_{max}+\frac{d}{dk}\delta_{\ell}\left(k\right)\right)
-\frac{1}{\pi k}\sin\left(kr_{max}-\ell\pi/2+\delta_{\ell}\left(k\right)
\right)\cos\left(kr_{max}-\ell\pi/2+\delta_{\ell}\left(k\right)\right) \\
=\frac{1}{\pi}\left\{r_{max}+\frac{d}{dk}\delta_{\ell}\left(k\right)
-\frac{1}{2 k}\sin\left[2\left(kr_{max}-\ell\pi/2+\delta_{\ell}
\left(k\right)\right)\right]\right\} \ .
\end{array}
\label{A6}
\end{equation}
 If we subtract from ($\ref{A6}$) the corresponding expression that would
be obtained if there were no interactions, we obtain for the oscillatory
terms\[
-\frac{(-1)^{\ell}}{ 2 \pi k}\left\{ \sin\left(2kr_{max}\right)\left(
\cos\left[2\delta_{\ell}\left(k\right)\right]-1\right)+
\cos\left(2kr_{max}\right)\sin\left[2\delta_{\ell}
\left(k\right)\right]\right\} \]
which has to be integrated over $k$, with a Boltzmann factor. Consequently,
what we wish to estimate are integrals such as 

\[
\begin{array}{c}
I_{1} =  \int _{0}^{\infty }\end{array}
dk\, \sin \left[2kr_{max}\right] \left(
\frac{\cos \left[2\delta_{\ell} \left(k\right)
\right] - 1}{k} \right) \, e^{- \beta \frac{\hbar^2}{m}k^2} \]
and\[
\begin{array}{c}
I_{2}= \int _{0}^{\infty }\end{array}
dk\, \cos \left(2kr_{max}\right)\frac{\sin \left[2\delta_{\ell}\left(k\right)
\right]}{k}\, e^{- \beta \frac{\hbar^2}{m}k^2} .           \]

In the vicinity of zero
 $\delta_{\ell}\left(k\right)=
\left(n+\epsilon/2\right)$$\pi+O\left(k^{2\ell+1}\right)$,
where $n$ is the number of bound states
supported by the potential and $\epsilon$ is zero except when $\ell=0$
and there is an s-wave zero-energy resonance, in which case $\epsilon$
equals 1.
Evaluating the integrals, either directly, or making the change of variable
$u = 2 k r_{max}$, and letting $r_{max}$ become large, we then find that there
are no contributions to the virial from the oscillatory terms in our
three dimensional case, unless there is a zero-energy resonance at $l = 0$,
in which case the phase shift
starts at $\pi/2$ and $I_{1}$ equals -$\pi$. The total contribution for 
$l = 0$, to be added to $1/\pi$ times the derivative of the phase shift, is
then $1/2$.

\section*{The One Dimensional Case}

In this section we discuss the 2nd-virial of a Bose one-dimensional
system of particles with repulsive $\delta$-function interactions.

The symmetric wave function for the 2-body system \cite{gaudin} in Cartesian
coordinates is
\begin{equation}
\begin{array}{c}
\psi_{k_{1},k_{2}}^{s}(x_{1},x_{2})=
\frac{1}{\sqrt{2!}}\left(\frac{1}{2\pi}\right)\frac{|k|}{\sqrt{k^{2}+
\mathit{g}^{2}}}\,\left[\left(1-\frac{i\mathit{g}}{k}\right)
e^{i\left(k_{1}x_{1}+k_{2}x_{2}\right)}+\left(1+\frac{i\mathit{g}}
{k}\right)e^{i\left(k_{2}x_{1}+k_{1}x_{2}\right)}\right]
\end{array}
\label{wfcc}
\end{equation}
where  $x_{i}$ is the
position coordinate of particle $i$ and $k_{1}$ and
$k_{2}$ are the two momenta (wave numbers) involved in the system.
The expression is only valid for $x_{2} > x_{1}$, but is correct for all 
momenta. The normalization is such that integrating over these values of 
$x_{i}$ gives the expected delta function normalization, as is appropriate
for a symmetric function.
The strength of the interaction is represented by $\mathrm{g}$; we define
$\mathit{g}=\left(m \mathrm{g}/2 \hbar^{2}\right)$.

We note that an expression \cite{amaya} valid for $x_{2} < x_{1}$, 
requires one  to interchange the coefficient of the exponentials 
in (\ref{wfcc}).

To derive an expression for the second virial coefficient of the one
dimensional system, we parallel the discussion
for the three dimensional case in the earlier sections.

We begin the discussion of the virial by rewriting 
$\left(\ref{wfcc}\right)$, and the wave functions in 
the second of Eqs. $\left(\ref{trace}\right)$,
in terms of center of mass and relative coordinates.
We then evaluate the C.M. part as in Eq. $\left(\ref{b2cm}\right)$ which,
 for a system contained on a line of length $A$,
equals
\begin{equation}
\begin{array}{rl}
\frac{1}{A}Tr^{s}\left(e^{-\beta T_{cm}}\right)= &
\frac{1}{2 \pi} \int_{-\infty}^{\infty} dK \,
 e^{-\left(\hbar^{2}/4m\right)\beta K^{2}} \\
= & {2^{1/2}}/{\lambda_{T}},
\end{array}
\label{1b2cm}
\end{equation}
 where $K=k_{1}+k_{2}$, and the associated position coordinate $X = (x_{1}
+ x_{2})/2$. 
\vspace{.1in}

The wave function for the relative motion is then
\begin{equation}
\psi_{k}^{s}(x)=  \frac{1}{\sqrt{2!}}(\frac{1}{2\pi})^{1/2}
\frac{|k|}{\sqrt{k^{2}+\mathit{g}^{2}}}\,\left[\left(
1-\frac{i\mathit{g}}{k}\right)e^{ikx}+\left(1+\frac{i\mathit{g}}{k}\right)
e^{-ikx}\right] ,
\label{1wf}
\end{equation}
where $k=\left(k_{2}-k_{1}\right)/2$ and the relative position is $x = x_{2}
- x_{1} > 0$.

A representation of this equation, valid for all values of $x$, and all
values of $k$,
can then be
 written as
\begin{equation}
\psi_{k}^{s}(x)    = (\frac{1}{\pi})^{1/2}\cos(k |x| + \delta(k)) ,
\end{equation}
in terms of the phase shift 
\begin{equation}
\delta(k)=-\frac{\pi}{2}  + \arctan\frac{k}{\mathit{g}},  \; \; \textnormal{for}
 \; k > 0, \; \; \textnormal{and} 
\; \;  \delta(-|k|) = - \delta(|k|) .
\label{ps}
\end{equation}
We note that our expression for the wave function is completely symmetric
in all of our variables! We have the choice of either integrating over
half of the $x$ space and the complete $k$ space, or over the complete
$x$ space and half of the $k$ space. We choose to do the latter.
For $k$ positive, we first integrate, over $x > 0$, the 
square of the wave function. We can do this by using our Green's function
`trick' as we have done previously for 3 dimensions, but since we know the
explicit form of the wave function in terms of its phase shift, it is
tempting to evaluate the integral directly!

\begin{equation}
\int_{0}^{x_{max}} \cos^{2}(kx + \delta(k))  dx = \frac{1}{2} x_{max} 
- \frac{1}{4 k} \sin(2 \delta(k)) + \frac{1}{4 k}\sin(2 k x_{max} + 
2\delta(k)),
\end{equation}
where we delay letting the upper limit $x_{max}$ go to infinity, until the
subtraction of the integral of the square of the `free' solution,
and the subsequent evaluation of the `oscillating' terms, when
integrating over $k$.
We don't immediately see the term in the derivative of the phase shift,
that was obtained from the limiting procedure in our former treatment,
but, we note from Eq. ($\ref{ps}$) that 
\[ -\frac{1}{2k} \sin(2 \delta(k)) = \frac{d}{dk} \delta(k) . \]
Integrating over 
$-x_{max}$ to $x_{max}$
(a simple factor of 2, due to
symmetry), and with the normalization, we obtain 

\begin{equation}
\begin{array}{c}
\int_{-x_{max}}^{x_{max}} (\frac{1}{\pi}\cos^{2}(k|x| + \delta(k)))  dx - 
\int_{-x_{max}}^{x_{max}} (\frac{1}{\pi}\cos^{2}(kx))  dx = \\
\frac{1}{\pi}\frac{d}{dk}\delta(k) +
\frac{1}{2 \pi k}\,[\sin(2 k x_{max})(\cos(2 \delta(k)) - 1) +
\cos(2 k x_{max}) \sin(2 \delta(k))] \ .
\end{array}
\end{equation}

We recognize the form of the oscillating terms, which apart from a change
of sign, is precisely that encountered in three dimensions.

Following our previous comments on the domain of the integrations, 
we integrate, over $k$, from $0$ to $\infty$, and evaluate
$I_{1}$ and $I_{2}$, using the phase shift given by ($\ref{ps}$). This gives 
$I_{1} = - \pi/2 $, $I_{2} = 0$.
Finally, joining the contributions of the oscillatory and non-oscillatory 
terms, we achieve the expression we were looking for, for the difference
of the fugacity coefficients in terms of asymptotic quantities,

\begin{equation}
b_{2}-b_{2}^{0}=
\frac{2^{1/2}}{\lambda_{T}}\left
[ - \frac{1}{2} +\int_{0}^{\infty}dk\, e^{-\left(\hbar^{2}/m\right)\beta k^{2}}
\frac{1}{\pi} \frac{d\delta}
{dk}(k) \right] \ .
\label{dif:}
\end{equation}

\section*{Equivalence of the Results of Dodd and Gibbs.}

In this section we show that our formalism, and $\left(\ref{dif:}\right)$,
 gives us an equivalent result
for the difference in the fugacity coefficients, to that 
obtained in ref. \cite{dodd}. While the former was derived
directly from the partition function and is expressed in terms of
asymptotic quantities, the latter was derived using the usual cluster
expansion \cite{lee} and is expressed in terms of integrals. \\
The derivative of the phase shift reads 
\begin{equation}
\frac{d\delta}{dk}=\frac{\mathit{g}}{k^{2}+\mathit{g}^{2}}
\label{dps}
\end{equation}
and, therefore, the 
integral of the phase shift term in Eq. ($\ref{dif:}$) is
\begin{equation}
\begin{array}{c}
\left(\frac{1}{\pi}\right)\frac{2^{1/2}}{\lambda_{T}}
\int_{0}^{\infty}dk\, e^{-\left(\hbar^{2}/m\right)\beta k^{2}}
\frac{\mathit{g}}{k^{2}+\mathit{g}^{2}}\\
=\frac{2^{1/2}}{\lambda_{T}}(\frac{1}{2})
e^{\left(\hbar^{2}/m\right)\beta\mathit{g}^{2}}\left(1-\textrm{erf}
\left[\mathit{g}\sqrt{\left(\hbar^{2}/m\right)\beta}\right]\right) \ .
\end{array}
\label{difb}
\end{equation}
Joining the contributions of the oscillatory and non-oscillatory 
terms, we obtain\begin{equation}
b_{2}-b_{2}^{0}=\frac{1}{2^{1/2}\lambda_{T}}\left[-1+
e^{\left(\hbar^{2}/m\right)\beta\mathit{g}^{2}}
\left(1-\textrm{erf}\left[\mathit{g}\sqrt{\left(\hbar^{2}/m\right)
\beta}\right]\right)\right].
\label{fin0}
\end{equation}
 This is exactly the result from Eq.(13) of ref. \cite{dodd},\[
\begin{array}{rl}
b_{2}-b_{2}^{0}= & \frac{1}{2\pi\left(\hbar^{2}/2m\right)\beta}
\int_{0}^{\infty}ds\,\,\exp\left[-s^{2}/\left(\left(\hbar^{2}/m\right)
\beta\right)\right]\left[\exp\left(-2\mathit{g}s\right)-1\right] 
,\end{array}
\]
when integrated over the variable $s$ .

\newpage
\section*{Acknowledgments}
A. Amaya acknowledges the CONACyT for partial support under project 41072-F,
and S. Y. Larsen thanks the Centro and the Institute for their 
hospitality and support.


\vspace{.5in}

\begin{thebibliography}{10}
\bibitem{dodd}L. R. Dodd and A. M. Gibbs, J. Math. Phys. \textbf{15,} 41 (1974)
\bibitem{hussien} N. A. R. Hussien and A. A. Yahya, J. Phys. A: Math.
Gen. {\bf 30}, 445 (1997)
\bibitem{dashen}D. Dashen, S. Ma and M.J. Bernstein, Phys. Rev.
 \textbf{187}, 345 (1969)
\bibitem{masche}S.Y. Larsen and P.L. Mascheroni, Phys. Rev. \textbf{A2},
 1018 (1970)
\bibitem{larsen2}S.Y. Larsen, A. Palma and M. Berrondo, J. Chem. Phys.
 \textbf{77}, 5816 (1982)
\bibitem{jei}S.Y. Larsen and J. Zhen, Mol. Phys. \textbf{63}, 581 (1988)
\bibitem{larsen}S. Larsen, Proceedings of the Bogolyubov Conference on
 Problems of Theoretical and Mathematical Physics, 1999.
 $\;$ physics$/0105074$\\
 Phys. Elem. Part. and Atom. Nucl., Part. and Nucl. \,{} \textbf{31},
\, 7B, 156 (2000)
\bibitem{gibson}W. Gibson, S.Y. Larsen, and J. Popiel, Phys Rev.
 \textbf{35}, 4919 (1987)
\bibitem{popiel}J.J. Popiel and S.Y. Larsen, Few-Body Systems
 \textbf{15}, 129 (1993)
\bibitem{amaya}A. Amaya-Tapia, S.Y. Larsen, and J. Popiel, Few-Body Systems
 \textbf{23}, 87 (1997)
\bibitem{gibson2}W.G. Gibson, Phys. Rev. \textbf{A36}, 564 (1987)
\bibitem{poll}S.Y. Larsen and J.D. Poll, Can. J. Phys, \textbf{52}, 1914
 (1974) 
\bibitem{gaudin}M. Gaudin, J. Math. Phys. \textbf{12} , 1674 (1971);
 J. Math. Phys.
\textbf{12}, 1677 (1971)
\bibitem{lee}T. D. Lee and C. N. Yang, Phys. Rev. \textbf{113}, 1165 (1959)
\bibitem{boyd}M. E. Boyd, S. Y. Larsen and J.E. Kilpatrick, J. Chem. Phys.
\textbf{45}, 499 (1966)
\bibitem{uhlenbeck}G. E. Uhlenbeck and E. Beth, Physica \textbf{3},
 729 (1936); E. Beth and G. E. Uhlenbeck, \emph{ibid}, 915 (1937)
\end{thebibliography}
\end{document}